\documentclass{article}
\usepackage{spconf,amsmath,graphicx,subcaption}
\usepackage[stable]{footmisc}
\usepackage[T1]{fontenc} 
\usepackage[utf8]{inputenc} 
\usepackage{amsmath,cite,url}
\usepackage{color}
\usepackage{multirow}

\title{Automatic Lyrics Alignment and Transcription in Polyphonic Music:\\Does Background music help?}
\name{Chitralekha Gupta, Emre Y{\i}lmaz, Haizhou Li}
\address{Department of Electrical and Computer Engineering, National University of Singapore}
\let\OLDthebibliography\thebibliography
\renewcommand\thebibliography[1]{
  \OLDthebibliography{#1}
  \setlength{\parskip}{0pt}
  \setlength{\itemsep}{0pt plus 0.3ex}
}
\begin{document}
\ninept
\maketitle
\begin{abstract}
Background music affects lyrics intelligibility of singing vocals in a music piece. Automatic lyrics alignment and transcription in polyphonic music are challenging tasks because the singing vocals are corrupted by the background music. In this work, we propose to learn music genre-specific characteristics to train polyphonic acoustic models. We first compare several automatic speech recognition pipelines for the application of lyrics transcription. We then present the lyrics alignment and transcription performance of music-informed acoustic models for the best-performing pipeline, and systematically study the impact of music genre and language model on the performance. With such genre-based approach, we explicitly model the music without removing it during acoustic modeling. The proposed approach outperforms all competing systems in the lyrics alignment and transcription tasks on several well-known polyphonic test datasets.
\end{abstract}
\vspace{-0.3cm}
\section{Introduction}
\label{sec:intro}
\vspace{-0.15cm}
Lyrics is an important component of music, and people often recognize a song by its lyrics. Lyrics contribute to the mood of the song \cite{ali2006songs}, affect the opinion of a listener about the song \cite{anderson1981love}, and even help in foreign language learning \cite{good2015efficacy}. Automatic lyrics alignment is the task of finding word boundaries of the given lyrics with the polyphonic audio, while transcription is the task of recognizing the sung lyrics from audio. These are useful for various music information retrieval applications such as generating karaoke scrolling lyrics, music video subtitling, query-by-singing \cite{hosoya2005lyrics}, keyword spotting, and automatic indexing of music according to transcribed keywords \cite{fujihara2008hyperlinking}.

Automatic lyrics transcription of singing vocals in the presence of background music remains an unsolved problem. One of the earliest studies \cite{gruhne2007phoneme} conducted frame-wise phoneme classification in polyphonic music where it was attempted to recognize three broad-classes of phonemes in 37 popular songs using acoustic features such as MFCCs, and PLP. Mesaros et al.~\cite{mesaros2010automatic} adopted an automatic speech recognition (ASR) based approach for phoneme and word recognition of singing vocals in monophonic and polyphonic music.
 
Singing vocals are often highly correlated with the corresponding background music, resulting in overlapping frequency components \cite{ramona2008}. To suppress the background accompaniment, many approaches have incorporated singing voice separation techniques as a pre-processing step \cite{mesaros2010automatic,gupta2019,fujihara2011lyricsynchronizer}. However, this step makes the system dependent on the performance of the singing voice separation algorithm, as the separation artifacts may make the words unrecognizable. Moreover, this requires a separate training setup for the singing voice separation system. In our latest work~\cite{gupta2019acoustic}, we trained acoustic models on a large amount of solo singing vocals and adapted them towards polyphonic music using a small amount of in-domain data -- extracted singing vocals, and polyphonic audio. We found that domain adaptation with polyphonic data outperforms that with extracted singing vocals. This suggests that acoustic model adapted with polyphonic data captures the spectro-temporal variations of vocals+background music better than that adapted with extracted singing vocals which have distortions and artifacts.

Recently, Stoller et al.~\cite{stoller2019} presented a data intensive end-to-end approach to lyrics transcription and alignment from raw polyphonic audio. However, end-to-end systems require a large amount of annotated training polyphonic music data to perform well, as seen in \cite{stoller2019} that uses more than 44,000 songs with line-level lyrics annotations from Spotify's proprietary music library, while publicly available resources for polyphonic music are limited.

Instead of treating the music as background noise, we hypothesize that acoustic models induced with music knowledge will help in lyrics alignment and transcription in polyphonic music. Schultz and Huron \cite{condit2015catching} found that genre-specific musical attributes such as instrumental accompaniment, singing vocal loudness, syllable rate, reverberation, and singing style influence human intelligibility of lyrics. In this study, we train genre-informed acoustic models for automatic lyrics transcription and alignment using an openly available polyphonic audio resource. We discuss several variations for the ASR components, such as, the acoustic model, and the language model (LM), and systematically study their impact on lyrics recognition and alignment accuracy on well-known polyphonic test datasets.
\vspace{-0.3cm}
\section{Lyrics alignment and transcription framework}
\vspace{-0.2cm}
Our goal is to build a designated ASR framework for automatic lyrics alignment and transcription. We explore and compare various approaches to understand the impact of background music, and the genre of the music on acoustic modeling. We detail the design procedure followed to gauge the impact of different factors in the following subsections. 
\vspace{-0.3cm}
\subsection{Singing vocal extraction vs.~polyphonic audio}
\label{sec:vocal_poly}
\vspace{-0.15cm}
Earlier approaches to lyrics transcription have used acoustic models that were trained on solo-singing audio. Singing vocal extraction was then applied on the test data \cite{mesaros2010automatic,gupta2019,dzhambazov2017knowledge}. Such acoustic models can be adapted to a small set of extracted vocals to reduce the mismatch of acoustic models between training and testing \cite{gupta2019acoustic}. Now that we have available a relatively large polyphonic lyrics annotated dataset (DALI)\cite{meseguer2018dali}, we explore two approaches for acoustic modeling for the task of lyrics transcription and alignment: (1) to apply singing vocal extraction from the polyphonic audio as a pre-processing step, and train acoustic models with the extracted singing vocals, and (2) to train acoustic models using the lyrics annotated polyphonic dataset directly. Approach (1) treats the background music as the background noise and suppresses it. On the other hand, approach (2) observes the combined effect of vocals and music on acoustic modeling. With these two approaches, we would like to answer the question whether background music helps in acoustic modeling for lyrics transcription and alignment.
\vspace{-0.3cm}
\subsection{Standard ASR vs.~end-to-end ASR}
\vspace{-0.15cm}
Given the state-of-the-art lyrics alignment and transcription system is an end-to-end ASR trained on a large polyphonic audio dataset \cite{stoller2019}, we compare the performance of a standard ASR pipeline, comprising of a separate acoustic model, language model, and pronunciation lexicon, with an end-to-end ASR on the lyrics transcription task using the limited publicly available resources. 

Hosoya et al.~\cite{hosoya2005lyrics} described lyrics recognition grammar using a finite state automaton (FSA) built from the lyrics in the queried database, so as to exploit the linguistic constraints in lyrics such as rhyming patterns, connecting words, and grammar \cite{fang2017discourse}. However, these methods have been tested only on small solo-singing datasets, and their scalability to a larger vocabulary recognition of polyphonic songs needs to be tested. In this work, we investigate the performance of standard N-gram techniques, also used for large vocabulary ASR, for lyrics transcription in polyphonic songs. We train two N-gram models: (1) an in-domain LM (henceforth referred to as the lyrics LM) trained only on the lyrics from the training music data and (2) a general LM trained on a large publicly available text corpus extracted from different resources. 

The lyrics transcription quality of the standard ASR architecture is compared with an end-to-end system, both trained on the same polyphonic training data. The end-to-end ASR approach learns how to map spectral audio features to characters without explicitly using a language model and pronunciation lexicon~\cite{graves2014,chan2016,watanabe2017,stoller2019}. The end-to-end ASR system is trained using a multiobjective learning framework with a connectionist temporal classification (CTC) objective function and an attention decoder appended to a shared encoder~\cite{watanabe2017}. A joint decoding scheme has been used to combine the information provided by the hybrid model consisting of CTC and attention decoder components and hypothesize the most likely recognition output.
\vspace{-0.3cm}
\subsection{Genre-informed acoustic modeling}
\vspace{-0.15cm}
Genre of a music piece is characterized by background instrumentation, rhythmic structure, and harmonic content of the music \cite{tzanetakis2002musical}. Factors such as instrumental accompaniment, vocal harmonization, and reverberation are expected to interfere with lyric intelligibility, while predictable rhyme schemes and semantic context might improve intelligibility \cite{condit2015catching}. They found that across 12 different genres, the overall lyrics intelligibility for humans is 71.7\% (i.e.~the percentage of correctly identified words from a total of 25,408 words), where ``Death Metal'' excerpts received intelligibility scores of zero, while ``Pop'' excerpts achieved scores close to 100\%.
\vspace{-0.3cm}
\subsubsection{Genre-informed phone models}
\begin{figure}[!t]
  \centering
  \begin{subfigure}{\linewidth}
    \centering
    \includegraphics[width=1\columnwidth,height=1cm]{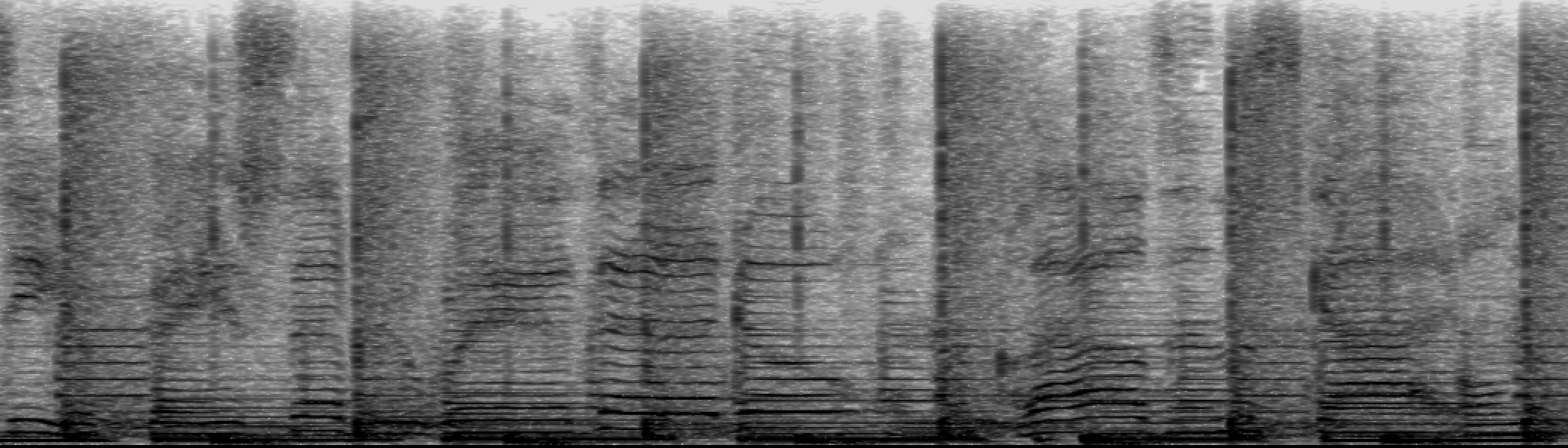}
    \vspace{-0.5cm}
    \caption{}
  \end{subfigure}
  \begin{subfigure}{\linewidth}
    \centering
    \includegraphics[width=1\columnwidth,height=1cm]{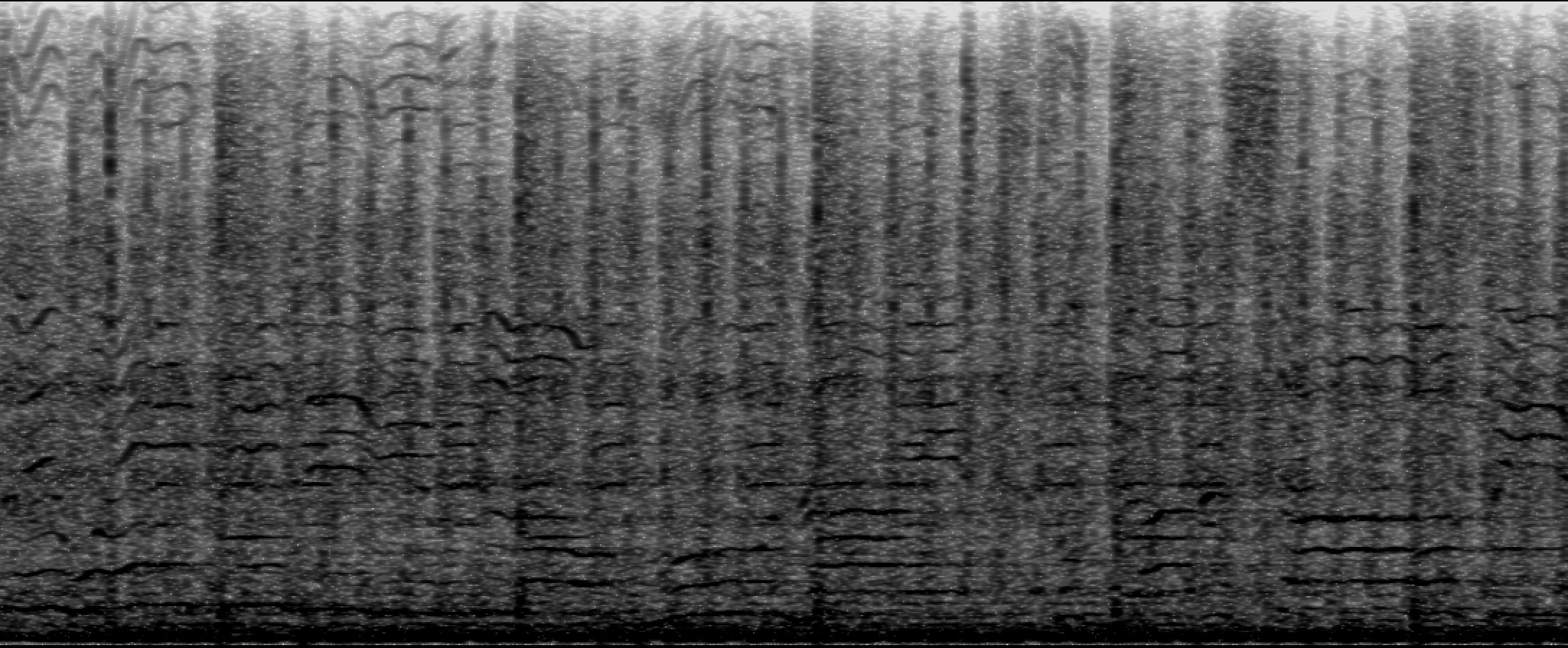}
    \vspace{-0.5cm}
    \caption{}
  \end{subfigure}
  \begin{subfigure}{\linewidth}
    \centering
    \includegraphics[width=1\columnwidth,height=1cm]{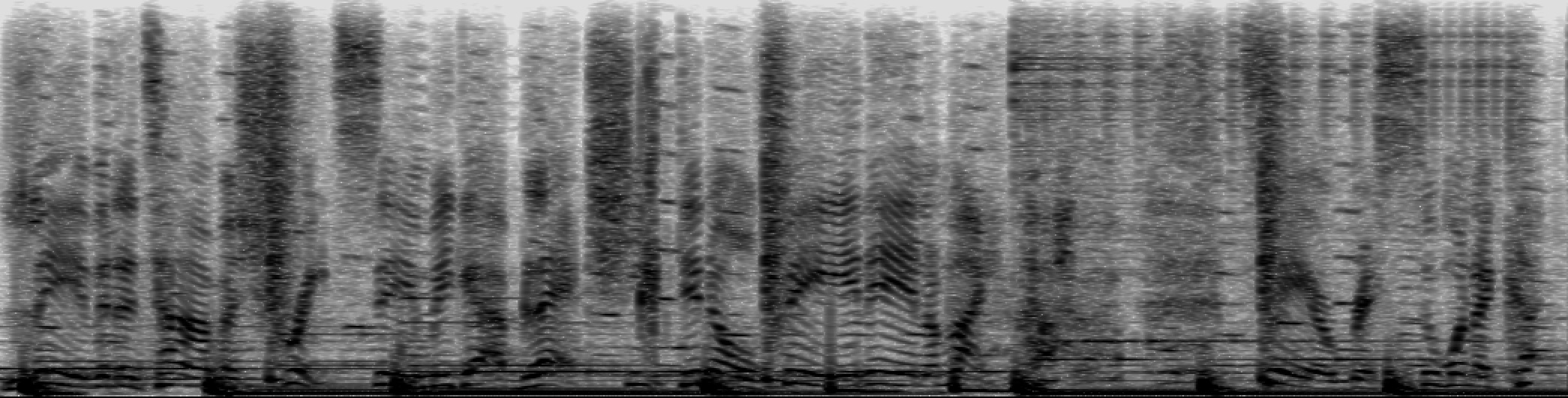}
    \vspace{-0.5cm}
    \caption{}
  \end{subfigure} 
  \vspace{-0.3cm}
  \caption{Spectrogram of 5 seconds audio clip of vocals with background music (sampling frequency:16kHz, spectrogram window size:64ms) for (a) Genre: \textit{Pop}; Song: \textit{Like the Sun}, by \textit{Explosive Ear Candy} (timestamps: 00:16-00:21) (b) Genre: \textit{Metal}; Song: \textit{Voices}, by \textit{The Rinn} (timestamps: 01:07-01:12), and (c) Genre: \textit{Hip Hop}; Song: \textit{The Statement}, by \textit{Wordsmith} (timestamps: 00:16-00:21).}
  \vspace{-0.1cm}
  \label{fig:specgram}
\end{figure}

\vspace{-0.15cm}
One main difference between genres that affects lyric intelligibility is the relative volume of the singing vocals compared to the background accompaniment. For example, as observed in \cite{condit2015catching}, in \textit{metal} songs, the accompaniment is loud and interferes with the vocals, while is relatively softer in \textit{jazz}, \textit{country}, and \textit{pop} songs. Figure \ref{fig:specgram}(a) is the spectrogram of a pop song excerpt showing loud singing vocals with visible singing voice harmonics. On the other hand, Figure \ref{fig:specgram}(b) shows the dense spectrogram of a metal song that has amplified distortion on electric guitar, and loud beats, with relatively soft singing vocals. Another difference between genres is the syllable rate. In \cite{condit2015catching}, it was observed that \textit{rap} songs, that have a higher syllable rate, show lower lyric intelligibility than other genres. The hip hop song in Figure \ref{fig:specgram}(c) has clear and rapid vocalization corresponding to a rhythmic speech in presence of beats. We believe that genre-specific acoustic modelling of phones would capture the combined effect of background music and singing vocals, depending on the genre, and help in automatic lyrics transcription and alignment.

\vspace{-0.4cm}
\subsubsection{Genre-informed ``silence'' models}
\vspace{-0.15cm}
In speech, there are long-duration non-vocal segments that include silence, background noise, and breathing. In an ASR system, a silence acoustic model is separately modeled for better alignment and recognition. Non-vocal segments or musical interludes are also frequently occurring in songs, especially between verses. However, in polyphonic songs, these non-vocal segments consist of different kinds of musical accompaniments that differ across genres. For example, a metal song typically consists of a mix of highly amplified distortion guitar, and emphatic percussive instruments, a typical jazz song consists of saxophone and piano, and a pop song consists of guitar and drums. The spectro-temporal characteristics of the combination of instruments vary across genres, but are somewhat similar within a genre. Thus, we propose to train genre-specific non-vocal or ``silence'' models to characterize this variability of instrumentation across genres. 
\vspace{-0.3cm}
\subsubsection{Genre broadclasses}
\vspace{-0.15cm}
\begin{table}[]
\caption{Genre broadclasses grouping}
\label{tab:genreclasses}
\vspace{-0.3cm}
\resizebox{\columnwidth}{!}{%
\begin{tabular}{|c|l|l|}
\hline
\textbf{\begin{tabular}[c]{@{}c@{}}Genre\\ Broadclasses\end{tabular}} & \multicolumn{1}{c|}{\textbf{Characteristics}} & \multicolumn{1}{c|}{\textbf{Genres}} \\ \hline
hiphop & rap, electronic music & Rap, Hip Hop, R\&B \\ \hline
metal & \begin{tabular}[c]{@{}l@{}}loud and many background \\accompaniments, a mix of percussive\\instruments, amplified distortion, vocals\\not very loud, rock, psychedelic\end{tabular} & \begin{tabular}[c]{@{}l@{}}Metal, Hard Rock,\\Electro, Alternative, \\Dance, Disco,\\Rock, Indie\end{tabular} \\ \hline
pop & \begin{tabular}[c]{@{}l@{}}vocals louder than the background\\ accompaniments, guitar, piano,\\saxophone, percussive instruments\end{tabular} & \begin{tabular}[c]{@{}l@{}}Country, Pop, Jazz,\\ Soul, Reggae, Blues,\\ Classical\end{tabular} \\ \hline
\end{tabular}%
\vspace{-1cm}
}
\end{table}
Music has been divided into different genres in many different and overlapping ways, based on a shared set of characteristics \cite{tzanetakis2002musical}.

To build genre-informed acoustic models, we consider the shared characteristics between genres that affect lyrics intelligibility, such as type of background accompaniments, and loudness of vocals, and group all genres to three broad genre classes: \textit{pop}, \textit{hiphop}, and \textit{metal}. Table \ref{tab:genreclasses} summarizes our genre broadclasses. We categorize songs containing some rap along with electronic music under \textit{hiphop} broadclass, which includes genres such as Rap, Hip Hop, and Rhythms \& Blues. Songs with loud and dense background music are categorized as \textit{metal}, that includes genres such as Metal and Hard Rock. Songs with clear and louder vocals under genres Pop, Country, Jazz, Reggae etc. are categorized as \textit{pop} broadclass.
\vspace{-0.3cm}
\section{Experimental Setup}
\vspace{-0.2cm}
We conduct three sets of experiments to demonstrate and compare different strategies for lyrics alignment and transcription: (1) train acoustic models using (a) extracted vocal and (b) polyphonic audio and compare their ASR performance, (2) compare a standard ASR system to an end-to-end ASR both trained on polyphonic music audio, and (3) compare the performance of genre-informed acoustic models to the genre-agnostic models, and also explore the impact of lyrics LM and general LM.
\vspace{-0.4cm}
\subsection{Datasets}
\label{ssec:4.1}
\vspace{-0.15cm}
All datasets used in the experiments are summarized in Table \ref{tab:datasets}. The training data for acoustic modeling contains 3,913 audio tracks.\footnote{Out of a total of 5,358 audio tracks in DALI, only 3,913 were English language and audio links were accessible from Singapore.} English polyphonic songs from the DALI dataset \cite{meseguer2018dali}, consisting of 180,033 lyrics-transcribed lines with a total duration of 134.5 hours.

We evaluated the performance of lyrics alignment and transcription on three test datasets - Hansen's polyphonic songs dataset (9 songs) \cite{hansen2012recognition}\footnote{The manual word boundaries of 2 songs in this dataset - \textit{clocks} and \textit{i kissed a girl} were not accurate, thus excluded them from the alignment study}, Mauch's dataset (20 songs) \cite{mauch2012integrating}, and Jamendo dataset (20 songs) \cite{stoller2019}. Hansen's and Mauch's datasets were used in the MIREX lyrics alignment challenges of 2017 and 2018. These datasets consist mainly of Western pop songs with manually annotated word-level transcription and boundaries. The Jamendo dataset consists of English songs from diverse genres, along with their lyrics transcription and manual word boundary annotations.
\begin{table}[]
\centering
\caption{Dataset description}
\label{tab:datasets}
\vspace{-0.3cm}
\resizebox{\columnwidth}{!}{%
\setlength\tabcolsep{1.5pt}
\begin{tabular}{|c|c|c|c|}
\hline
\textbf{Name} & \textbf{Content} & \textbf{Lyrics Ground-Truth} & \textbf{Genre distribution}\\ \hline
\multicolumn{4}{|c|}{\textbf{Training data}} \\ \hline
DALI \cite{meseguer2018dali} & \begin{tabular}[c]{@{}c@{}}3,913\\songs\end{tabular}& \begin{tabular}[c]{@{}c@{}}line-level boundaries,\\180,033 lines\end{tabular} &\begin{tabular}[c]{@{}c@{}}hiphop:119,\\metal:1,576, pop:2,218\end{tabular}\\ \hline 
\multicolumn{4}{|c|}{\textbf{Test data}} \\ \hline
Hansen \cite{hansen2012recognition} & 9 songs & word-level boundaries, 2,212 words & hiphop:1, metal:3, pop:5\\ 
Mauch \cite{mauch2012integrating} & 20 songs & word-level boundaries, 5,052 words &hiphop:0, metal:8, pop:12\\
Jamendo \cite{stoller2019}&20 songs&word-level boundaries, 5,677 words &hiphop:4, metal:7, pop:9\\\hline
\end{tabular}%
}
\vspace{-0.6cm}
\end{table}

The genre tags for most of the songs in the training dataset (DALI) is provided in their metadata, except for 840 songs. For these songs, we applied an automatic genre recognition implementation \cite{genreclassifier} which has 80\% classification accuracy, to get their genre tags. We applied the genre groupings from Table \ref{tab:genreclasses} to assign a genre broadclass to every song. For the songs in the test datasets, we scanned the web to find their genre tags and categorized them into the three genre broadclasses. The distribution of the number of songs across the three genre broadclasses for all the datasets is shown in Table \ref{tab:datasets}. This distribution in the training data is skewed towards \textit{pop}, while \textit{hiphop} is the most under-represented. However, we are limited by the amount of data available for training, with DALI being the only resource. Therefore, we assume this to be the naturally occurring distribution of songs across genres.
\vspace{-0.3cm}
\subsection{Vocal separated data vs.~polyphonic data}
\vspace{-0.15cm}
As discussed in Section \ref{sec:vocal_poly}, we compare the strategies of vocal extracted data vs.~polyphonic data to train the acoustic models, as a way to find out if the presence of background music helps. We use the reported best performing models M4, from the state-of-the-art Wave-U-Net based audio source separation algorithm \cite{stoller2018,githubvocalsep} for separating vocals from the polyphonic audio. 
\vspace{-0.3cm}
\subsection{ASR framework: standard ASR vs.~end-to-end ASR}
\vspace{-0.15cm}
The ASR system used in these experiments is trained using the Kaldi ASR toolkit~\cite{povey2011kaldi}. A factorized time-delay neural network (TDNN-F) model~\cite{povey2018} with additional convolutional layers (2 convolutional, 10 time-delay layers followed by a rank reduction layer) was trained according to the standard Kaldi recipe (version 5.4) using 40-dimensional MFCCs as acoustic features of an augmented version of the polyphonic training data (Section~\ref{ssec:4.1})~\cite{ko2015}. The default hyperparameters provided in the standard recipe were used and no hyperparameter tuning was performed during the acoustic model training. A duration-based modified pronunciation lexicon is employed which is detailed in~\cite{gupta2018automatic}. Two language models are trained using the transcriptions of the in-domain song-lyrics of DALI dataset (Lyrics LM) and the open source text corpus\footnote{http://www.openslr.org/11/} released as a part of the Librispeech corpus~\cite{librispeech} (general LM).

The end-to-end system is trained using the ESPnet toolkit \cite{watanabe2018espnet}. The shared encoder is a combination of two VGG~\cite{simonyan2014deep} layers followed by a BLSTM with subsampling~\cite{chan2016} with 5 layers and 1024 units. The attention-based decoder is a 2-layer decoder with 1024 units with coverage attention~\cite{See_2017}. The batchsize is set to 20 to avoid GPU memory overflow. The rest of the hyperparameters are consistent with the standard Librispeech recipe available in the toolkit (version 0.3.1). In pilot experiments, using a language model during the decoding with the default language model weight provided worse results than decoding without a language model. Therefore, no LM is used during the decoding step to avoid parameter tuning on the test data.
\vspace{-0.4cm}
\subsection{Genre-informed acoustic modeling}
\vspace{-0.15cm}
We train 3 different types of acoustic models corresponding to the three genre broadclasses, for (a) genre-informed ``silence'' or non-vocal models and (b) genre-informed phone models. We extract the non-vocal segments at the start and the end of each line in the training data for the training of ``silence'' model. For the genre-informed phone modeling, we label the phone units in the phonetic lexicon with genre labels. For the alignment task, we use the same genre-informed phone models that are mapped to the words without genre tags, i.e.~the alignment system chooses the best fitting phone models among all genres during the forced alignment, to prevent the additional requirement of genre information for songs in the test sets.

\vspace{-0.3cm}
\section{Results and Discussion~\footnote{D\lowercase{emo: \texttt{https://lyrics-demo.droppages.com/}}}}
\vspace{-0.2cm}
\subsection{Singing vocal extraction vs.~polyphonic audio}
\vspace{-0.15cm}
We compare the performance of a standard ASR trained on extracted singing vocals and polyphonic audio for the tasks of lyrics alignment (Table \ref{tab:res_ali}) and transcription (Table \ref{tab:vocext_poly}).
The alignment performance is measured as the mean absolute word boundary error (AE) for each song, averaged over all songs of a dataset, in seconds \cite{stoller2019,dzhambazov2017knowledge}, and lyrics transcription performance is measured as the word error rate (WER) which is a standard performance measure for ASR systems. We see an improvement in both alignment and transcription performance with ASR trained on polyphonic data than vocal extracted data, on all the test datasets. This indicates that there is value in modeling the combination of vocals and music, instead of considering the background music as noise and suppressing it. Although we have used the state-of-the-art vocal extraction algorithm, these techniques are still not perfect, and introduce artifacts and distortions in the extracted vocals, which is the reason for poor performance of the models trained with extracted vocals. AE has reduced to less than 350 ms in all the test datasets using polyphonic models given in the third column of Table \ref{tab:res_ali}. We observe a large improvement in the alignment accuracy on the Mauch's dataset. It consists of many songs with long musical interludes, where the extracted vocals models fail to align the lyrics around the long non-vocal sections because of erratic music suppression. Polyphonic models, on the other hand, are able to capture the transitions from music to singing vocals. In the following experiments, we use polyphonic audio to train the acoustic models.

\begin{table}[]
\caption{Mean absolute word alignment error (AE)(seconds)}
\label{tab:res_ali}
\vspace{-0.2cm}
\resizebox{\columnwidth}{!}{%
\begin{tabular}{|c|c|c|c|c|}
\hline
\textbf{\begin{tabular}[c]{@{}c@{}}Test\\ Datasets\end{tabular}} & \textbf{\begin{tabular}[c]{@{}c@{}}Vocal\\ Extracted\end{tabular}} & \textbf{\begin{tabular}[c]{@{}c@{}}Polyphonic:\\ No Genre Info\end{tabular}} & \textbf{\begin{tabular}[c]{@{}c@{}}Polyphonic\\ Genre Silence\end{tabular}} & \textbf{\begin{tabular}[c]{@{}c@{}}Polyphonic:Genre\\ Silence+Phone\end{tabular}} \\ \hline
Mauch & 3.62 & 0.25 & 0.28 & \textbf{0.21} \\ \hline
Hansen & 0.67 & \textbf{0.16} & 0.25 & 0.18 \\ \hline
Jamendo & 0.39 & 0.34 & 0.42 & \textbf{0.22} \\ \hline
\end{tabular}%
}
\vspace{-0.3cm}
\end{table}
\begin{table}[t]
\caption{Lyrics transcription WER (\%) comparison of vocal extracted vs. polyphonic data trained acoustic models}
\label{tab:vocext_poly}
\vspace{-0.2cm}
\centering
\resizebox{0.7\columnwidth}{!}{%
\begin{tabular}{|c|c|c|}
\hline
\multicolumn{1}{|l|}{\textbf{Test datasets}} & \multicolumn{1}{l|}{\textbf{Vocal extracted}} & \multicolumn{1}{l|}{\textbf{Polyphonic}} \\ \hline
Mauch & 76.31 & 54.08 \\ \hline
Hansen & 78.85 & 60.77 \\ \hline
Jamendo & 71.83 & 66.58 \\ \hline
\end{tabular}%
}
\vspace{-0.4cm}
\end{table}
\vspace{-0.4cm}
\subsection{Standard ASR vs.~end-to-end ASR}
\vspace{-0.15cm}
The end-to-end ASR's lyrics transcription performance reported in Table \ref{tab:std_vs_end2end} is comparable to the Stoller's end-to-end system \cite{stoller2019}, which was however trained on a much larger dataset. The standard ASR performs considerably better than the end-to-end ASR, as can be seen in the second column of Table \ref{tab:std_vs_end2end}. This implies that characterizing different components of polyphonic music with the standard ASR components using acoustic model, pronunciation model, and language model are valuable for the task of lyrics transcription. The following experiments use the standard ASR framework for exploring genre-informed acoustic modeling.
\begin{table}[]
\caption{Comparison of lyrics transcription WER (\%) of Standard ASR vs.~End-to-end ASR}
\label{tab:std_vs_end2end}
\vspace{-0.2cm}
\centering
\resizebox{0.6\columnwidth}{!}{%
\begin{tabular}{|c|c|c|}
\hline
\textbf{Test Datasets} & \textbf{Standard} & \textbf{End-to-end} \\ \hline
Mauch & 54.08 & 73.2 \\ \hline
Hansen & 60.77 & 80.1 \\ \hline
Jamendo & 66.58 & 87.9 \\ \hline
\end{tabular}%
}
\vspace{-0.5cm}
\end{table}
\vspace{-0.4cm}
\subsection{Genre-informed acoustic modeling}
\begin{figure}[t!]
    \centering
    \begin{subfigure}[t]{0.5\columnwidth}
        \centering
        \includegraphics[width=\columnwidth,height=2.8cm]{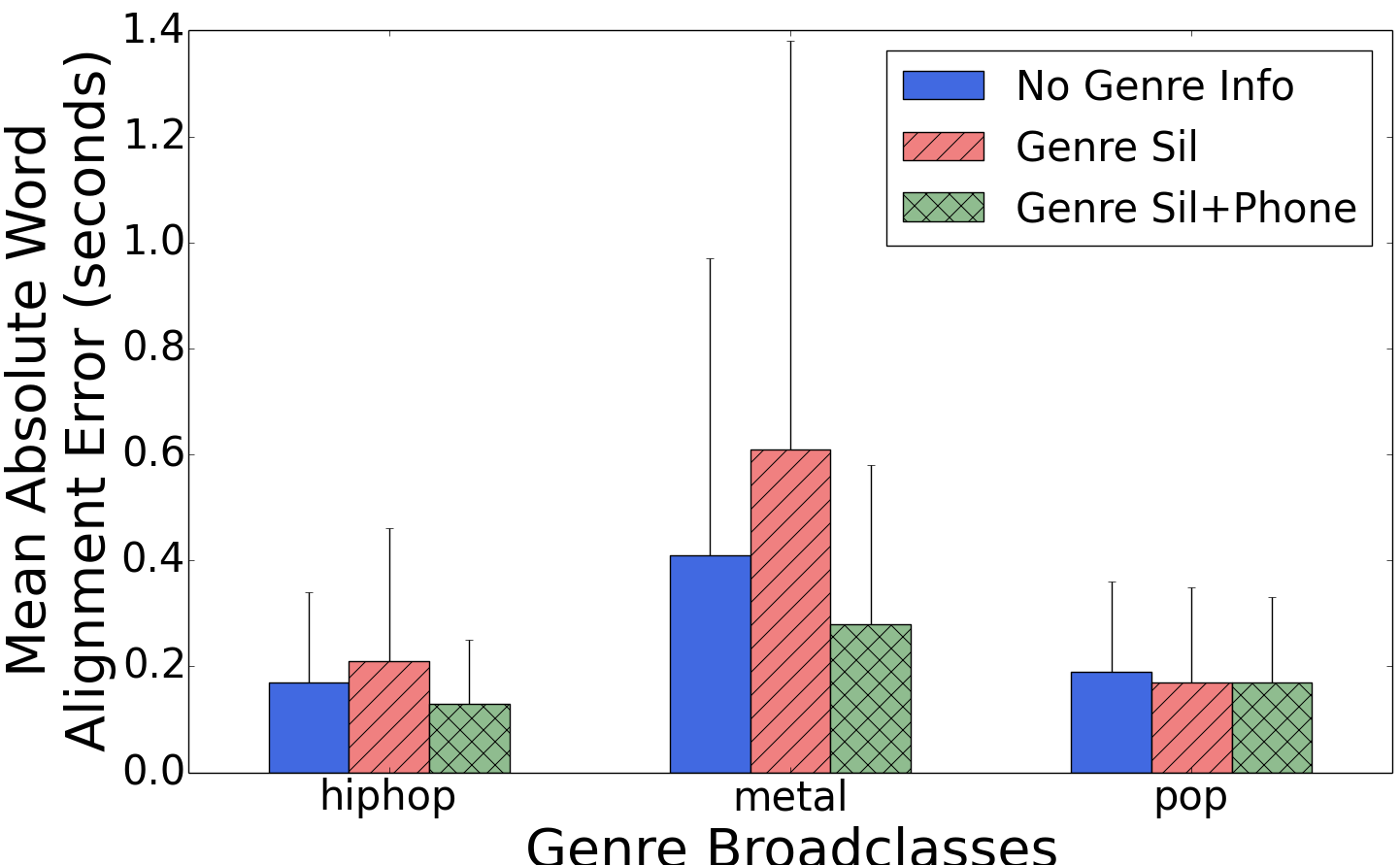}
        \vspace{-0.3cm}
        \caption{}
    \end{subfigure}
    \hspace{-0.5em}
    \begin{subfigure}[t]{0.5\columnwidth}
        \centering  \includegraphics[width=\columnwidth]{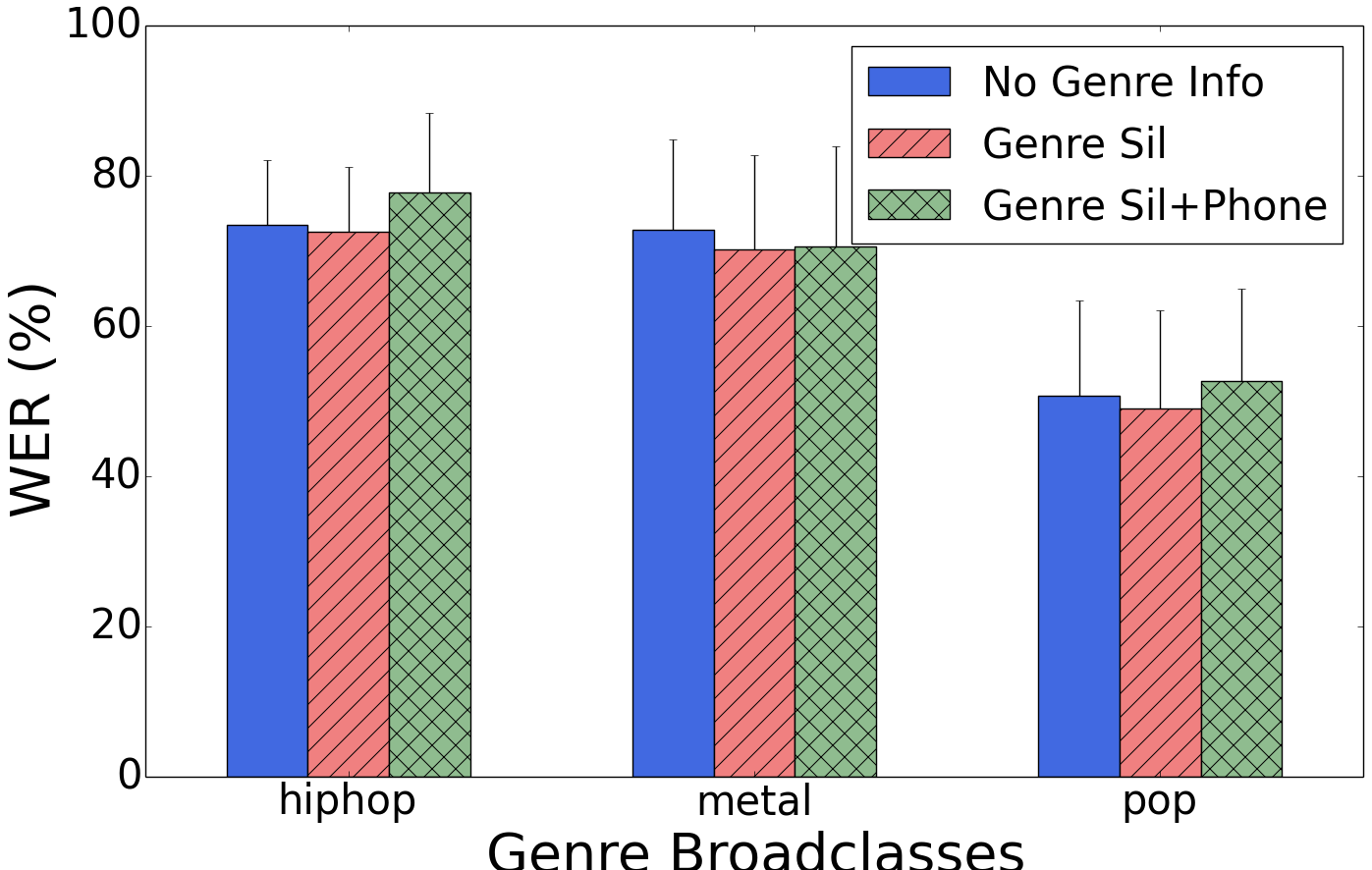}
        \vspace{-0.3cm}
        \caption{}
    \end{subfigure}%
    \vspace{-0.4cm}
    \caption{Comparison of (a) lyrics alignment AE (seconds), and (b) lyrics transcription WER (\%) across all the test datasets.}
    \label{fig:genre_analysis}
    \vspace{-0.3cm}
\end{figure}

\vspace{-0.15cm}
Lyrics alignment shows an improvement in performance with genre-informed silence+phone models compared to genre-agnostic (or no genre info) and genre-informed silence models, as seen in Table \ref{tab:res_ali}. AE is less than 220 ms across all test datasets. This indicates that for the task of lyrics alignment where the target lyrics are known, the genre-informed phone models trained on limited data are able to capture the transition between phones well. Figure \ref{fig:genre_analysis}(a) shows that the alignment error is maximum in metal songs, which is intuitive due to the loud noisy background music.

The lyrics transcription performance for genre-informed \textit{silence} and \textit{silence+phone} models using two kinds of LM are presented in Table \ref{tab:genre_results}. The genre-informed silence models show 2-4\% absolute improvement in the word error rate (WER) over the genre-agnostic models in all the test datasets. This indicates that creating genre-specific models for the non-vocal segments is a good strategy to capture the variability of music across genres. However, genre-informed phone models do not show any improvement in WER. This could be due to the insufficient amount of data to train accurate phone models for three genre types. Hiphop class has the least amount of data, while pop has the most. Figure \ref{fig:genre_analysis}(a) indicates that the performance degradation is more in hiphop songs, than in pop songs. The performance on the metal songs improves with genre-informed silence models, however the WER is high despite the class having data comparable to that for pop songs. This suggests that the loud and dense background music in metal genre hinders the process of learning the singing vocal characteristics for accurate lyrics transcription.

Additionally, we observe an improvement in the lyrics transcription performance with the lyrics LM over general LM. This shows that the linguistic constraints due to the rhyming structure of the lyrics, are better captured by in-domain (song-lyrics) text, rather than by a general text collected from various textual resources.
\begin{table}[]
\caption{Comparison of lyrics transcription WER (\%)}
\label{tab:genre_results}
\vspace{-0.2cm}
\centering
\resizebox{\columnwidth}{!}{%
\begin{tabular}{|c|c|c|c|}
\hline
\textbf{Test Datasets} & \textbf{\begin{tabular}[c]{@{}c@{}}No Genre Info\end{tabular}} & \textbf{\begin{tabular}[c]{@{}c@{}}Genre Silence\end{tabular}} & \textbf{\begin{tabular}[c]{@{}c@{}}Genre Silence+Phone\end{tabular}} \\ \hline
\multicolumn{4}{|c|}{\textbf{General LM}} \\ \hline
Mauch & 54.08 & 52.45 & 53.74 \\ \hline
Hansen & 60.77 & 59.10 & 62.71 \\ \hline
Jamendo & 66.58 & 64.42 & 67.91 \\ \hline
\multicolumn{4}{|c|}{\textbf{Lyrics LM}} \\ \hline
Mauch & 45.78 & 44.02 & 45.70 \\ \hline
Hansen & 50.35 & 47.01 & 51.32 \\ \hline
Jamendo & 62.64 & 59.57 & 61.90 \\ \hline
\end{tabular}%
}
\vspace{-0.3cm}
\end{table}
\vspace{-0.4cm}
\subsection{Comparison with existing literature}
\vspace{-0.15cm}
In Table \ref{tab:litreview}, we compare our best results with the most recent prior work. Our strategy provides the best results for both lyrics alignment and transcription tasks on several datasets. The proposed strategies show a way to induce music knowledge in ASR to address the problem of lyrics alignment and transcription in polyphonic audio.
\begin{table}[!t]
\centering
\caption{Comparison of lyrics alignment (AE (seconds)) and transcription (WER\%) performance with existing literature.}
\label{tab:litreview}
\vspace{-0.2cm}
\resizebox{\columnwidth}{!}{%
\setlength\tabcolsep{1.5pt}
\begin{tabular}{|c|c|c|c|c|c|c|c|}
\hline
\textbf{} & \multicolumn{2}{c|}{\textbf{MIREX 2017}} & \textbf{MIREX 2018} & \multicolumn{2}{c|}{\textbf{ICASSP 2019}} & \textbf{Interspeech2019} & \textbf{} \\ \hline
\textbf{} & \textbf{AK\cite{kruspe2016bootstrapping}} & \textbf{GD\cite{dzhambazov2015modeling,dzhambazov2017knowledge}} & \textbf{CW \cite{wang2018mirex}} & \textbf{DS\cite{stoller2019}} & \textbf{CG\cite{gupta2019}} & \textbf{CG\cite{gupta2019acoustic}} & \textbf{Ours} \\ \hline
\multicolumn{8}{|c|}{\textbf{Lyrics Alignment}}\\\hline
\textbf{Mauch} & 9.03 & 11.64 & 4.13 & 0.35 & 6.34 & 1.93 & 0.21\\ 
\textbf{Hansen} & 7.34 & 10.57 & 2.07 & - & 1.39 & 0.93 & 0.18\\ 
\textbf{Jamendo} & - & - & - & 0.82 & - & - & 0.22\\ \hline
\multicolumn{8}{|c|}{\textbf{Lyrics Transcription}}\\\hline
\textbf{Mauch} & - & - & - & 70.9 & - & - & 44.0\\ 
\textbf{Hansen} & - & - & - & - & - & - & 47.0\\ 
\textbf{Jamendo} & - & - & - & 77.8 & - & - & 59.6\\ \hline
\end{tabular}%
}
\vspace{-0.5cm}
\end{table}
\vspace{-0.4cm}
\section{Conclusions}
\vspace{-0.2cm}
In this work, we introduce a music-informed strategy to train polyphonic acoustic models for the tasks of lyrics alignment and transcription in polyphonic music. We model the genre-specific characteristics of music and vocals, and study their performance with different ASR frameworks, and language models. We find that this music-informed strategy learns the background music characteristics that affect lyrics intelligibility, and shows improvement in lyrics alignment and transcription performance over others with music suppression. We also show that with limited available data, our strategy of genre-informed acoustic modeling as well as lyrics constrained language modeling in a standard ASR pipeline is able to outperform all existing systems for both lyrics alignment and transcription tasks.

\bibliographystyle{IEEEbib}
\bibliography{strings,refs_EY}

\begin{thebibliography}{10}

\bibitem{ali2006songs}
S.~O. Ali and Z.~F. Peynircio{\u{g}}lu,
\newblock ``Songs and emotions: are lyrics and melodies equal partners?,''
\newblock {\em Psychology of Music}, vol. 34, no. 4, pp. 511--534, 2006.

\bibitem{anderson1981love}
B.~Anderson, D.~Berger, R.~Denisoff, K.~Etzkorn, and P.~Hesbacher,
\newblock ``Love negative lyrics: Some shifts in stature and alterations in
  song,''
\newblock {\em Communications}, vol. 7, no. 1, pp. 3--20, 1981.

\bibitem{good2015efficacy}
A.~J. Good, F.~A. Russo, and J.~Sullivan,
\newblock ``The efficacy of singing in foreign-language learning,''
\newblock {\em Psychology of Music}, vol. 43, no. 5, pp. 627--640, 2015.

\bibitem{hosoya2005lyrics}
T.~Hosoya, M.~Suzuki, A.~Ito, S.~Makino, L.~A. Smith, D.~Bainbridge, and I.~H.
  Witten,
\newblock ``Lyrics recognition from a singing voice based on finite state
  automaton for music information retrieval.,''
\newblock in {\em ISMIR}, 2005, pp. 532--535.

\bibitem{fujihara2008hyperlinking}
H.~Fujihara, M.~Goto, and J.~Ogata,
\newblock ``Hyperlinking lyrics: A method for creating hyperlinks between
  phrases in song lyrics.,''
\newblock in {\em ISMIR}, 2008, pp. 281--286.

\bibitem{gruhne2007phoneme}
M.~Gruhne, C.~Dittmar, and K.~Schmidt,
\newblock ``Phoneme recognition in popular music.,''
\newblock in {\em ISMIR}, 2007, pp. 369--370.

\bibitem{mesaros2010automatic}
A.~Mesaros and T.~Virtanen,
\newblock ``Automatic recognition of lyrics in singing,''
\newblock {\em EURASIP Journal on Audio, Speech, and Music Processing}, vol.
  2010, no. 1, pp. 546047, 2010.

\bibitem{ramona2008}
M.~Ramona, G.~Richard, and B.~David,
\newblock ``Vocal detection in music with support vector machines,''
\newblock in {\em 2008 Proc. ICASSP}. IEEE, 2008, pp. 1885--1888.

\bibitem{gupta2019}
C.\* Gupta, B.\* Sharma, H.~Li, and Y.~Wang,
\newblock ``Automatic lyrics-to-audio alignment on polyphonic music using
  singing-adapted acoustic models,''
\newblock in {\em Proc. ICASSP}. IEEE, 2019, pp. 396--400.

\bibitem{fujihara2011lyricsynchronizer}
M.~Fujihara, H.and~Goto, J.~Ogata, and H.~G. Okuno,
\newblock ``Lyricsynchronizer: Automatic synchronization system between musical
  audio signals and lyrics,''
\newblock {\em IEEE Journal of Selected Topics in Signal Processing}, vol. 5,
  no. 6, pp. 1252--1261, 2011.

\bibitem{gupta2019acoustic}
C.~Gupta, E.~Y{\i}lmaz, and H.~Li,
\newblock ``Acoustic modeling for automatic lyrics-to-audio alignment,''
\newblock in {\em Proc. INTERSPEECH}, Sept. 2019, pp. 2040--2044.

\bibitem{stoller2019}
D.~Stoller, S.~Durand, and S.~Ewert,
\newblock ``End-to-end lyrics alignment for polyphonic music using an
  audio-to-character recognition model,''
\newblock in {\em Proc. ICASSP}. IEEE, 2019, pp. 181--185.

\bibitem{condit2015catching}
N.~Condit-Schultz and D.~Huron,
\newblock ``Catching the lyrics: intelligibility in twelve song genres,''
\newblock {\em Music Perception: An Interdisciplinary Journal}, vol. 32, no. 5,
  pp. 470--483, 2015.

\bibitem{dzhambazov2017knowledge}
G.~Dzhambazov,
\newblock {\em Knowledge-based Probabilistic Modeling for Tracking Lyrics in
  Music Audio Signals},
\newblock Ph.D. thesis, Universitat Pompeu Fabra, 2017.

\bibitem{meseguer2018dali}
G.~Meseguer-Brocal, A.~Cohen-Hadria, and G.~Peeters,
\newblock ``Dali: A large dataset of synchronized audio, lyrics and notes,
  automatically created using teacher-student machine learning paradigm,''
\newblock in {\em Proc. ISMIR}, 2018.

\bibitem{fang2017discourse}
J.~Fang, D.~Grunberg, D.~T. Litman, and Y.~Wang,
\newblock ``Discourse analysis of lyric and lyric-based classification of
  music.,''
\newblock in {\em ISMIR}, 2017, pp. 464--471.

\bibitem{graves2014}
A.~Graves and N.~Jaitly,
\newblock ``Towards end-to-end speech recognition with recurrent neural
  networks,''
\newblock in {\em Proceedings of the 31st International Conference on
  International Conference on Machine Learning - Volume 32}. 2014, ICML'14, pp.
  II--1764--II--1772, JMLR.org.

\bibitem{chan2016}
W.~{Chan}, N.~{Jaitly}, Q.~{Le}, and O.~{Vinyals},
\newblock ``Listen, attend and spell: A neural network for large vocabulary
  conversational speech recognition,''
\newblock in {\em Proc. ICASSP)}, March 2016, pp. 4960--4964.

\bibitem{watanabe2017}
S.~{Watanabe}, T.~{Hori}, S.~{Kim}, J.~R. {Hershey}, and T.~{Hayashi},
\newblock ``Hybrid {CTC/Attention} architecture for end-to-end speech
  recognition,''
\newblock {\em IEEE Journal of Selected Topics in Signal Processing}, vol. 11,
  no. 8, pp. 1240--1253, Dec 2017.

\bibitem{tzanetakis2002musical}
G.~Tzanetakis and P.~Cook,
\newblock ``Musical genre classification of audio signals,''
\newblock {\em IEEE Transactions on Speech and Audio Processing}, vol. 10, no.
  5, pp. 293--302, 2002.

\bibitem{hansen2012recognition}
J.~K. Hansen,
\newblock ``Recognition of phonemes in a-cappella recordings using temporal
  patterns and mel frequency cepstral coefficients,''
\newblock in {\em 9th Sound and Music Computing Conference (SMC)}, 2012, pp.
  494--499.

\bibitem{mauch2012integrating}
M.~Mauch, H.~Fujihara, and M.~Goto,
\newblock ``Integrating additional chord information into hmm-based
  lyrics-to-audio alignment,''
\newblock {\em IEEE Transactions on Audio, Speech, and Language Processing},
  vol. 20, no. 1, pp. 200--210, 2012.

\bibitem{genreclassifier}
``Musical genre recognition using a cnn,''
  \url{https://github.com/thomas-bouvier/music-genre-recognition.git},
\newblock [Online; accessed 5-July-2019].

\bibitem{stoller2018}
D.~Stoller, S.~Ewert, and S.~Dixon,
\newblock ``Wave-u-net: A multi-scale neural network for end-to-end audio
  source separation,''
\newblock in {\em Proc. ISMIR}, 2018.

\bibitem{githubvocalsep}
``Implementation of the wave-u-net for audio source separation,''
  \url{https://github.com/f90/Wave-U-Net.git},
\newblock [Online; accessed 5-July-2019].

\bibitem{povey2011kaldi}
A.~Povey, D.and~Ghoshal, G.~Boulianne, L.~Burget, O.~Glembek, N.~Goel,
  M.~Hannemann, P.~Motlicek, Y.~Qian, P.~Schwarz, J.~Silovsky, G.~Stemmer, and
  K.~Vesely,
\newblock ``The {Kaldi} speech recognition toolkit,''
\newblock in {\em in Proc. ASRU}, 2011.

\bibitem{povey2018}
D.~Povey, G.~Cheng, Y.~Wang, K.~Li, H.~Xu, M.~Yarmohammadi, and S.~Khudanpur,
\newblock ``Semi-orthogonal low-rank matrix factorization for deep neural
  networks,''
\newblock in {\em Proc. INTERSPEECH}, 2018, pp. 3743--3747.

\bibitem{ko2015}
T.~Ko, V.~Peddinti, D.~Povey, and S.~Khudanpur,
\newblock ``Audio augmentation for speech recognition,''
\newblock in {\em Proc. INTERSPEECH}, 2015, pp. 3586--3589.

\bibitem{gupta2018automatic}
C.~Gupta, H.~Li, and Y.~Wang,
\newblock ``Automatic pronunciation evaluation of singing,''
\newblock {\em Proc. INTERSPEECH}, pp. 1507--1511, 2018.

\bibitem{librispeech}
V.~{Panayotov}, G.~{Chen}, D.~{Povey}, and S.~{Khudanpur},
\newblock ``Librispeech: An {ASR} corpus based on public domain audio books,''
\newblock in {\em Proc. ICASSP}, April 2015, pp. 5206--5210.

\bibitem{watanabe2018espnet}
S.~Watanabe, T.~Hori, S.~Karita, T.~Hayashi, J.~Nishitoba, Y.~Unno, N.~Enrique
  Yalta~Soplin, J.~Heymann, M.~Wiesner, N.~Chen, A.~Renduchintala, and
  T.~Ochiai,
\newblock ``Espnet: End-to-end speech processing toolkit,''
\newblock in {\em Proc. INTERSPEECH}, 2018, pp. 2207--2211.

\bibitem{simonyan2014deep}
K.~Simonyan and A.~Zisserman,
\newblock ``Very deep convolutional networks for large-scale image
  recognition,'' 2014.

\bibitem{See_2017}
A.~See, P.~J. Liu, and C.~D. Manning,
\newblock ``Get to the point: Summarization with pointer-generator networks,''
\newblock {\em Proceedings of the 55th Annual Meeting of the Association for
  Computational Linguistics (Volume 1: Long Papers)}, 2017.

\bibitem{kruspe2016bootstrapping}
A.~M. Kruspe,
\newblock ``Bootstrapping a system for phoneme recognition and keyword spotting
  in unaccompanied singing.,''
\newblock in {\em ISMIR}, 2016, pp. 358--364.

\bibitem{dzhambazov2015modeling}
G.~B. Dzhambazov and X.~Serra,
\newblock ``Modeling of phoneme durations for alignment between polyphonic
  audio and lyrics,''
\newblock in {\em 12th Sound and Music Computing Conference}, 2015, pp.
  281--286.

\bibitem{wang2018mirex}
Chung-Che Wang,
\newblock ``Mirex2018: Lyrics-to-audio alignment for instrument accompanied
  singings,''
\newblock in {\em MIREX 2018}, 2018.

\end{thebibliography}

\end{document}